\documentclass{article}
\usepackage[english]{babel}

\usepackage[letterpaper,top=2cm,bottom=2cm,left=3cm,right=3cm,marginparwidth=1.75cm]{geometry}

\usepackage{amsmath}
\usepackage{graphicx}
\usepackage[colorlinks=true, allcolors=blue]{hyperref}
\usepackage{amssymb}
\usepackage{amsthm}
\usepackage{todonotes}
\usepackage{xspace}
\usepackage{complexity}
\usepackage{authblk}

\newtheorem{proposition}{Proposition}
\newtheorem{definition}{Definition}
\newtheorem{theorem}{Theorem}
\newtheorem{example}{Example}
\newtheorem{lemma}{Lemma}
\newtheorem{corollary}{Corollary}
\newtheorem{claim}{Claim}

\newenvironment{cproof}
{\begin{proof}
 [Proof.]
 \vspace{-1.5\parsep}
}
{ \end{proof}}

\usepackage{cleveref}

\newcommand{\ignore}[1]{}

\newcommand{\costoutEmpty}{\mathrm{cost}}
\newcommand{\costoutwithrootEmpty}{\widehat{{\mathrm{cost}}}}
\newcommand{\costout}[1]{\costoutEmpty\left(#1\right)}
\newcommand{\costoutwithroot}[1]{\costoutwithrootEmpty\left(#1\right)}

\newcommand{\plan}{\tau}

\newcommand{\pacplan}{\plan_{\textsc{pac}}}
\newcommand{\halfplan}{\plan_{\textsc{half}}}

\newcommand{\s}{s}
\newcommand{\instance}{S}
\newcommand{\leaves}[1]{S\left(#1\right)}

\newcommand{\mutliplicationOrdering}{\textsc{MultOrd}\xspace}
\newcommand{\eval}[1]{|#1|}
\newcommand{\opt}[1]{\mathrm{OPT}\left(#1\right)}
\newcommand{\optwithroot}[1]{\widehat{\mathrm{OPT}}\left(#1\right)}

\newcommand{\uao}[3]{\underset{#1}{\overset{#2}{#3}}}

\newcommand{\ppb}{\textsc{ProdPart}\xspace}
\newcommand{\partition}{\textsc{2-Partition}\xspace}
\newcommand{\fnmax}[1]{m_{#1}}
\newcommand{\weight}[1]{w_{#1}}

\newcommand\set[1]{\{#1\}}
\newcommand{\join}{\bowtie}

\title{Approximation Algorithms for Product Query Optimization}
\author[1]{Macéo Ottavy}
\author[1]{Avijeet Ghosh} 
\author[1]{Pascal Koiran}
\author[1]{Alantha Newman}
\author[1]{\protect\\ François Schwarzentruber}
\author[2,3,1]{Blair D. Sullivan}
\author[1]{St\'ephan Thomassé}

\affil[1]{ENS de Lyon, CNRS, University Claude Bernard Lyon 1, LIP, UMR 5668, 69342, Lyon Cedex 07, France}
\affil[2]{University of Utah, Salt Lake City, Utah, USA}
\affil[3]{Collegium de Lyon, Lyon, France}

\begin{document}
\maketitle

\begin{abstract}

We study database query optimization for Cartesian products.
This problem is equivalent to finding an optimal plan for multiplying non-negative integers. We call this problem multiplication ordering. We show that the problem has an FPTAS when the cost of a plan is the sum of all intermediate products; both in the case when the overall final product is counted and when it is not. We show that using the optimal solution to the well-known product partition problem does not always lead to an optimal plan, but we
establish connections between the product partition and multiplication ordering problems. Finally, we prove that multiplication ordering is in PTIME (resp. \NP-complete) when integers are powers of a same base where the exponents are written in unary (resp. in binary).
\end{abstract}

\section{Introduction}
\label{section:introduction}

In database systems, query optimization consists in choosing the right actions to perform for computing the result of a query. For instance, should we compute the plan $((a \join b) \join c)$ or the plan $((b \join c) \join a)$, where $\join$ is the join operation for answering to the following SQL query?

\begin{verbatim}
SELECT a.*
  FROM a,
       b,
       c
 WHERE a.a_id = b.a_id
   AND b.b_id = c.b_id
\end{verbatim}

In a seminal paper, \cite{DBLP:journals/tods/IbarakiK84} proved that computing an optimal plan is \NP-complete; their plans are left-deep join trees and they model the cost by means of a selectivity factor between two relations $R_i$ and $R_j$ giving an estimation on the proportion of the size of $R_i \join R_j$ compared to $R_i \times R_j$. They ask to compute a plan that only uses the join operations explicitly given in the query.
In \cite{DBLP:conf/icdt/CluetM95}, \NP-hardness is proven for a more general notion of cost satisfying so-called ASI (adjacent sequence interchange) property
and allowing for computing any Cartesian products in the plan.
In \cite{DBLP:conf/pods/ScheufeleM97}, it was proven that the problem is also \NP-complete when plans are arbitrary trees (bushy trees) and only Cartesian products are considered. They consider the output cost that is the sum of the sizes (i.e., number of tuples) of all intermediate products computed during the plan.

\begin{example}
Consider relations $R_1, R_2, R_3, R_4$ of respective sizes 2, 2, 5, 10. The plan $(R_1 \times R_2) \times (R_3 \times R_4)$ can be represented by the following tree on which we have written the intermediate product sizes:

\newcommand\productsizeinfig[1]{\colorbox{gray!20!white}{#1}}
\begin{center}
\begin{tikzpicture}[
    level distance=10mm,
    level 1/.style={sibling distance=40mm},
    level 2/.style={sibling distance=20mm},
    edge from parent/.style={draw, thick},
    every node/.style={inner sep=0.5mm, font=\small}
]

\node {$\times$ \productsizeinfig{200}}
    child {
        node {$\times$ \productsizeinfig{10}}
        child { node {$R_3$ \productsizeinfig 5} }
        child { node {$R_2$ \productsizeinfig 2} }
    }
    child {
        node {$\times$ \productsizeinfig{20}}
        child { node {$R_4$ \productsizeinfig {10}} }
        child { node {$R_1$ \productsizeinfig 2} }
    };

\end{tikzpicture}
\end{center}

The output cost of this plan is $5+2+10+2+10+20+200 = 249$. As the final product size is always the same, we may also exclude it in the cost. The output cost without the root is $5+2+10+2+10+20 = 49$. 
\end{example}

The cost of a plan $\plan$ including the root is denoted by $\costoutwithroot{\plan}$ while its cost without the root is denoted by $\costout{\plan}$.
Although this query optimization problem is called cross product optimization in \cite{DBLP:conf/pods/ScheufeleM97}, we change its name to \mutliplicationOrdering to emphasize the fact that the input instance only contains the sizes of the relations. In particular the optimization process is agnostic about the very content of the relations. Thus, \mutliplicationOrdering corresponds to the natural problem of computing Cartesian products (i.e., joins without any conditions; no \texttt{WHERE} clauses in SQL) and where only the sizes of the relations are used to determine an optimal plan.

Observe that the \NP-hardness proven in \cite{DBLP:conf/pods/ScheufeleM97} relies on a reduction from Exact Covering With 3-Sets, in which elements of sets are encoded by relations where sizes are prime numbers. This setting is rather rare and one may wonder whether the problem is really so difficult in general. Note that the reduction in \cite{DBLP:conf/pods/ScheufeleM97}
involves the first prime numbers making the resulting instance of \mutliplicationOrdering in poly-size even if numbers are given in unary.
Said differently, \mutliplicationOrdering is NP-hard when numbers are given in unary. As the numbers represent sizes of relations written explicitly, it is practically relevant to consider the numbers written unary. 
In the current paper, we consider the numbers to be written binary making our approximation results stronger.

\subsection*{Our Contribution}

We show that \mutliplicationOrdering, although \NP-hard, is easy to approximate. This contrasts with the inaproximability of query optimization in general \cite{DBLP:conf/pods/ChatterjiEGY02}.

\begin{itemize}
    \item We show, see \Cref{thm:FPTAS with root for k>=2}, that \mutliplicationOrdering has an FPTAS (fully polynomial-time approximation scheme), meaning that we can find a $(1+\epsilon)$-approximate plan, in polytime in the size of the input and in $1/\epsilon$. We assume that the size of the relation sizes are given in binary.
    \item In the output cost, we always count the product of the sizes of all relations in the root. As this final product is present in all plans, we also consider the notion of cost in which we \emph{ignore the root} (i.e., we do not count the global product of all the inputs). In the previous example, the cost is then $5+2+10+2+10+20 = 49$. \Cref{thm:FPTASwithoutRoot} shows that \mutliplicationOrdering with the cost ignoring the global product at the root also has an FPTAS.
    
    \item We also compare multiplication ordering to product partition. First we show that an approximate solution for \ppb helps to find an approximate solution for \mutliplicationOrdering (\Cref{lemma:product partition is 2-APX}). However, applying product partition is not alway part of the optimal plan, and 
    we exhibit an example where applying a product partition at the root does not lead to an optimal plan for multiplication ordering (see \Cref{prop:productPartitionCounterExample}).
    
    \item Interestingly, we also focus on \mutliplicationOrdering when the sizes have a form of regularity (in contrast to being prime numbers as in the \NP-hardness proof \cite{DBLP:journals/tods/IbarakiK84}). We prove that when the sizes are all powers of the same base (e.g., $4^5, 4^{12}, 4^{12}, 4^{459}$, then  \mutliplicationOrdering is in PTIME (see \Cref{theorem:power}). However the problem becomes \NP-complete when the exponents are written in binary (see 
\Cref{theorem:powersexponentsinbinary}).
\end{itemize}

\paragraph{Paper outline.} 
In \Cref{section:relatedwork}, we briefly discuss the related work both in database theory and in algorithms. 
In \Cref{section:pb} we formally define \mutliplicationOrdering. 
Then, in \Cref{section:exactsolving}, we explain how to exactly solve \mutliplicationOrdering with a naïve algorithm; this is important because it gives subroutines for the approximation algorithms.
In \Cref{section:preprocessing}, we give two preprocessing rules allowing us to only consider input numbers greater or equal to $2$.
In \Cref{section:bounds}, we introduce two strategies: \textsc{Pacman}-plan and \textsc{Half}-plan and prove upper and lower bounds on the cost of any plan.
In \Cref{section:fptas}, we show that \mutliplicationOrdering has a FPTAS both for $\costoutwithrootEmpty$ and $\costoutEmpty$. 
In \Cref{section:linkToProdPart}, we highlight strong links between the famous \textsc{ProductPartition} problem and \mutliplicationOrdering.
In \Cref{section:restriction}, we then tackle the restriction on the input to powers. 
In \Cref{section:futurework}, we discuss future work.

\section{Related work}
\label{section:relatedwork}

The \mutliplicationOrdering problem has direct connections to problems in database query optimization as well as to ordering problems studied in other areas.

\subsection{Database theory}

Ordering of joins in database is a vast topic, see Table p. 45 in \cite{moerkotte2006building}. The problem is \NP-complete in general, see the seminal paper \cite{DBLP:journals/tods/IbarakiK84} in which they prove that it is \NP-hardness to get an optimal left-deep join tree where we also give a \emph{selectivity factor} for each pair of relations that can be joined. Later, in \cite{DBLP:conf/icdt/CluetM95}, they proved that this problem is also \NP-complete if we allow for Cartesian products, meaning that all pair of relations are joinable. 
In \cite{DBLP:conf/pods/ChatterjiEGY02}, it is shown that query optimization in general cannot be approximated.

The case when all selectivity factors are equal to 1 corresponds to computing Cartesian products.
In particular, searching for an optimal (with the output cost) bushy
\footnote{In the database theory literature, they (e.g., \cite{moerkotte2006building}) use the adjective `bushy' to emphasize that the tree can have an arbitrary shape.}
tree for computing a Cartesian product is \NP-hard \cite{DBLP:conf/pods/ScheufeleM97}. This problem is exactly \mutliplicationOrdering.
Observe that if we know the left-right order of the leaves of the tree, we can find OPT by dynamic programming. However nothing was known concerning approximation.

\subsection{Ordering in the context of other multiplication problems}

In algorithms, many problems involve ordering operations on a certain number of objects from some class.  Sometimes the outcome is always the same (e.g., multiplying integers or matrices), but intermediate objects of vastly different sizes can be created, depending on the ordering of operations.  Thus, an important goal in this context is to try to order these operations (e.g., multiplications) so as to reduce the sizes of the intermediate objects.

The most famous problem of this kind is finding the optimal ordering for the computation of a products of matrices, $M_1\cdot M_2\cdots M_n$.  This task can be done in $O(n^3)$ and is a standard textbook exercise in dynamic programming \cite{DBLP:books/daglib/0017733}.  Dynamic programming can be applied here because the left-right order of the matrices is fixed; only the optimal bushy tree has to be computed.

Many multiplication ordering problems are \NP-hard. One problem which seems close is the \textsc{VertexEliminationProblem} \cite{DBLP:journals/corr/abs-2506-17521} consisting of iteratively eliminating internal nodes in a directed graphs, namely adding edges from all in-neighbors to all out-neighbors and deleting the node. The cost to minimize is the sum of all in-degree times out-degree at the moment of the elimination of the node.
Another related work can be found in~\cite{abo2021approximation}.

\section{Definition of the {\normalfont \mutliplicationOrdering} problem}
\label{section:pb}

An instance of \mutliplicationOrdering is a multiset of positive integers $\instance = \{\s_i\}^n_{i=1}$.
A plan $\plan$ for computing $\s_1 \times \dots \times \s_n$ is a binary tree whose leaves are $\s_1, \dots, \s_n$.
A plan $\plan$ can be seen as a syntactic tree of an expression for computing $\s_1 \times \dots \times \s_n$ with parenthesis.
For the rest of the paper, unless stated otherwise, $n$ denotes the number of integers in the instance.  We sometimes refer to these input integers $\{s_1, \ldots, s_n\}$ as {\emph{elements}}.

\begin{example}
A plan for $2 \times 2 \times 5 \times 10$ could be $(5 \times 2) \times (10 \times 2)$ that corresponds to the syntactic tree \begin{tikzpicture}[
    level distance=5mm,
    level 1/.style={sibling distance=1cm},
    level 2/.style={sibling distance=5mm},
    edge from parent/.style={draw, thick},
    every node/.style={inner sep=0.5mm, font=\small}
]

\node {$\times$}
    child {
        node {$\times$}
        child { node {5} }
        child { node {2} }
    }
    child {
        node {$\times$}
        child { node {10} }
        child { node {2} }
    };

\end{tikzpicture}.
\end{example}

For a plan $\plan$, we denote by $\leaves{\plan}$ the multiset composed of the labels of the leaves of $\plan$. 
The evaluation of a plan $\plan$ is 
$|\plan| = \underset{\s\in\leaves{\plan}}{\prod} s$, 
that is the product of all leaves of $\plan$. The cost of a plan $\plan$ (denoted $\costout{\plan}$) is the sum of all the intermediate products at internal nodes of $\plan$ (we exclude the root and include the leaves). The cost that also includes the root is denoted by $\costoutwithroot{\plan}$. The formal definitions are as follows.

\begin{definition}[cost including the root]
For a leaf $s$, $\costoutwithroot{s} = s$.
For a plan $\plan_1 \times \plan_2$, $\costoutwithroot{\plan_1 \times \plan_2} := |\plan_1| \cdot |\plan_2| + \costoutwithroot{\plan_1} + \costoutwithroot{\plan_2}$.
\end{definition}

\begin{definition}[cost excluding the root]
For al plan $\plan$, $\costout{\plan} := \costoutwithroot{\plan} - |\plan|$.
\end{definition}

Alternatively, the cost excluding the root can also be defined by induction as follows. The base case is $\costout{\s} = 0$ for a leaf $\s$ which is also a root in that case.
The cost of the leaves will be accounted for during their evaluation. For a plan $\plan = \plan_1 \times \plan_2$, we have $\costout{\plan} = \costout{\plan_1} + \costout{\plan_2} + \eval{\plan_1} + \eval{\plan_2}$.

\begin{example}
If $\plan$ is $(5 \times 2) \times (10 \times 2)$ then $\costout{\plan} = 5+2+10+2+10+20 = 49$ and $\costoutwithroot{\plan} = 5+2+10+2+10+20+200 = 249$.
\end{example}

\begin{definition}\mutliplicationOrdering is the following optimization problem: given a multiset $\instance = \{\s_i\}^n_{i=1}$ of positive integers written in binary, compute a plan $\plan$ to compute $\s_1 \times \dots \times \s_n$ minimizing $\costout{\plan}$.
\end{definition}

We will also consider the variant \mutliplicationOrdering where we minimize $\costoutwithroot{\plan}$. To avoid ambiguity, we talk about `\mutliplicationOrdering for $\costoutEmpty$' and `\mutliplicationOrdering for $\costoutwithrootEmpty$'. Note that the distinction is not important for exact solving, since in this case, the two problems are equivalent. However the distinction is important when considering approximation ratios.

Let $\instance$ be an instance of \mutliplicationOrdering and $X\subset\instance$ a non-empty subset of $\instance$.  We say that a plan $\plan$ is {\emph{following $X$}} if it is of the form $\plan = \plan_1 \times\plan_2$ such that $\leaves{\plan_1} = X$ and $\leaves{\plan_2} = \instance\backslash X$. Moreover, for any subset $X\subseteq \instance$, we denote the \emph{weight} of $X$ in $\instance$ to be $\weight{\instance}(X) = \underset{\s\in X}{\prod} \s + \underset{\s\in\instance\backslash X}{\prod} \s$. 
Therefore the cost of a plan $\plan=\plan_1\times\plan_2$ following $X\subseteq\instance$ is $\costout{\plan} = \costout{\plan_1} + \costout{\plan_2} + \weight{\instance}(X)$.

\section{Exact solving}
\label{section:exactsolving}
For an instance $\instance$, we denote by $\opt{\instance}$ (resp. $\optwithroot{\instance}$) the optimal cost of the \mutliplicationOrdering problem for this instance using the $\costoutEmpty$ (resp. $\costoutwithrootEmpty$) objective function.

Solving \mutliplicationOrdering could be decomposed into phases: find the right permutation on the leaves and then choose the inner structure of the tree. The following proposition states that if the right permutation has been found, the problem can be solved in polynomial time. This is especially useful since some algorithms in databases rely on a particular choice of permutation. Given an instance $\instance = \{\s_i\}^n_{i=1}$ of \mutliplicationOrdering, a plan computing $\s_1\times \cdots\times \s_n$ respecting the order is a plan which in its associated bracketing, element $\s_i$ appears before $\s_j$ if and only if $i<j$. Equivalently, it is a plan in which the binary tree's leaves are $\s_1,\dots,\s_n$ in that order and such that no edges of the tree are crossing.
\begin{example}
For the instance $\{2,5,7,8\}$, the plan $(2\times 5) \times (7\times 8)$ computes the product $2\times 5\times 7\times 8$ and respects the order, but the plan $(2\times 8) \times (5\times 7)$ does not respect the order.
\end{example}
\begin{proposition}
Let $\{\s_i\}^n_{i=1}$ be an instance of \mutliplicationOrdering. One can find an optimal plan respecting the order in time polynomial in the size of the instance.
\label{proposition:polytimewithfixedpermutation}
\end{proposition}
\begin{proof}
We propose an algorithm based on dynamic programming. The subproblems are $M[i,j]$, which
denote the optimal value of $\costoutwithrootEmpty$ of a plan to compute $\s_i\times\dots\times\s_j$ that follows the order $\s_i, \dots, \s_j$. We now prove that:
$$\forall i\leq j,\; M[i,j] = \begin{cases}
\s_i & \text{ if } i=j,\\
\underset{i\leq a\leq j-1}{\min}\left\{M[i,a]+M[a+1,j] + \underset{b=i}{\overset{j}{\prod}}\s_b 
\right\} & \text{ otherwise.}
\end{cases}$$
The base case is true. The inductive case is proven as follows. In the sequel, we say that $\plan$ is a plan for $\s_i, \dots, \s_j$ when it is a plan for computing $\s_i\times \dots \times \s_j$ in that order.
Thus,
\begin{align*}
   M[i, j] =&\; \underset{\plan \text{ plan for $\s_i, \dots, \s_j$ }}{\min} \costoutwithroot{\plan}\\
   =&\underset{i\leq a\leq j-1}{\min}~
    \underset{\plan_2 \text{ plan for $\s_{a+1}, \dots, \s_j$ }}{\underset{
    \plan_1 \text{ plan for $\s_i, \dots, \s_a$ }}{\min}} ~
  \left(
        \costoutwithroot{\plan_1} + \costoutwithroot{\plan_2} + \eval{\plan_1} \times \eval{\plan_2} \right)\\
        =&\; \underset{i\leq a\leq j-1}{\min}\left\{M[i,a]+M[a+1,j] + \underset{b=i}{\overset{j}{\prod}}\s_b \right\}.
\end{align*}
We can design a dynamic programming algorithm that fills a table $M$ of size $\mathcal{O}(n^2)$. Each step costs at most $\mathcal{O}(n^2)$ arithmetic operations, resulting in a complexity of $\mathcal{O}(n^4)$. 
\end{proof}
Since we can enumerate all permutations of $\instance$ and solve the problem for each permutation using \Cref{proposition:polytimewithfixedpermutation}, the final complexity is  $n!\mathcal{O}(n^4) = \mathcal{O}(n!)$ since $n!\times n^4 \leq (n+4)!$.

\begin{corollary}
\label{cor:exactSolvingInn!}

\mutliplicationOrdering is solvable in $\mathcal{O}(n!)$.
\end{corollary}

We can actually obtain a faster bound.
\begin{theorem}
\label{thm:exact solving in 3^n}
\mutliplicationOrdering can be solved in $\mathcal{O}(3^n)$ time.
\end{theorem}
\begin{proof}
Let $\instance$ be an instance of \mutliplicationOrdering.  At each step, every plan splits the instance into two child subplans. Thus,
\begin{align*}
    \optwithroot{\instance} = \underset{\s\in\instance}{\prod}\s + \underset{\emptyset\neq X\subsetneq \instance}{\min} \left(\optwithroot{X} + \optwithroot{\instance\backslash X} \right).
\end{align*}
Then, consider a dynamic programming algorithm for this formula. It has to feed an array of size 
$|\{\emptyset\neq X\subsetneq \instance\}|=2^n-2$. At each step, it goes though all non-trivial subsets. We denote by $\s_{\max}$ the biggest $\s_i$. Notice that the cost of computing the product of some $\s_i$'s is equal to the sum of their representations' sizes. Therefore the computation cost for the instance $\instance$ is:
\begin{align*}
    \underset{\emptyset\neq X\subset \instance}{\sum} \left(
    |\{Y \mid \emptyset\neq Y\subset X\}| + |X| \right)
        \leq&\; \underset{\emptyset\neq X\subset\instance}{\sum} 2^{|X|}-2 + |X|\\
        \leq&\; \underset{i=0}{\overset{n}{\sum}} \binom{n}{i}\left(2^i + i)\right)\\
        \leq&\; 3^n + n2^n\\
        =&\; \mathcal{O}(3^n + n2^n).
\end{align*}
Using the binomial theorem stating that $\forall x,y,a\in\mathbb{N}, (x+y)^a = \uao{i=0}{a}{\sum} \binom{a}{i} x^iy^{n-i}$.  Thus the computation times is $\mathcal{O}(3^n)$. Moreover, note that minimizing $\costoutwithrootEmpty$ is the same as minimizing $\costoutEmpty$.
\end{proof}

\section{Preprocessing}
\label{section:preprocessing}
In this section, we highlight some safe preprocessing rules that can be applied at the beginning of any algorithm so that we can restrict our attention to more specific instances. 

\begin{proposition}
\label{prop:separationoftheinstance}
Let $\instance$ be an instance of \mutliplicationOrdering and let $\emptyset\neq \instance' \subsetneq\instance$, then $\opt{\instance} \leq \optwithroot{\instance'} + \optwithroot{\instance\backslash \instance'}$.
\end{proposition}
\begin{proof}
Let $\plan_1$ (resp. $\plan_2$) be an optimal plan for $\instance'$ (resp. $\instance\backslash \instance'$) for the $\costoutwithrootEmpty$ objective function. Let $\plan=\plan_1\times \plan_2$. Note that $\plan$ is a valid plan for the instance $\instance$ since $\instance = \instance' \sqcup (\instance\backslash \instance')$ where $\sqcup$ is the disjoint union operator.
Then $\opt{\instance} \leq \costout{\plan} = \costoutwithroot{\plan_1} + \costoutwithroot{\plan_2} = \optwithroot{\instance'} + \optwithroot{\instance\backslash \instance'}$.
\end{proof}

\begin{proposition}
\label{prop:someKEquaOne}
Let $\instance$ be an instance of \mutliplicationOrdering such that one input element is equal to $1$.  Then there is an optimal strategy that starts by multiplying this element with the next smallest element.
\end{proposition}
\begin{proof}
Assume $\instance=\{\s_1,\dots,\s_n\}$, that $s_1 = 1$ and that $\s_2 = \min\{\s_i,\; 1<i\leq n\}$ be the next smallest input element. Let $\plan^\ast$ be an optimal strategy on $\instance$ and $\plan_2$ be an optimal strategy on the instance $\{\s_2,\dots,\s_n\}$. Let $\plan$ be the strategy where we first multiply $\s_1$ and $\s_2$, then we apply the strategy of $\plan_2$. We claim that $\plan$ is an optimal plan. To prove it, first notice that $\costout{\plan}= \costout{\plan_2} +\s_1 +\s_2 = \s_1 +\s_2 + \opt{\instance\backslash\{s_1\}}$ since the multiplication of $\s_1$ and $\s_2$ is accounted as a leaf in $\plan_2$.
Now, observe that in $\plan^\ast$, $\s_1$ is multiplied with some subtree $\plan^*_2$. Let $\plan^\ast_1 = \s_1\times \plan^*_2$ be the subplan appearing in $\plan^\ast$ and $\instance'$ be the set of numbers involved in $\plan^\ast_2$. First note that $\costoutwithroot{\plan^\ast_1} =  \costoutwithroot{\plan^\ast_2} + \eval{\s_1} + \eval{\plan^\ast_1}  = \costoutwithroot{\plan_2^\ast} + \s_1 +\eval{\plan^\ast_2}$ and that $\eval{\plan^\ast_2}\geq \s_2$ by definition of $\s_2$. Then,
\begin{align*}
    \costout{\plan^\ast}
        \geq&\; \costoutwithroot{\plan^\ast_1} + \optwithroot{\instance\backslash (\instance'\cup\{\s_1\})}\\
        \geq&\; \costoutwithroot{\plan^\ast_2} + \s_1 + \s_2 + \optwithroot{\instance\backslash (\instance'\cup\{\s_1\})}\\
        \geq&\; \optwithroot{\instance'} + \optwithroot{\instance\backslash (\instance'\cup\{\s_1\})} + \s_1 + \s_2\\
        \geq&\; \s_1+\s_2 + \opt{\instance\backslash\{\s_1\}} \text{ (By \Cref{prop:separationoftheinstance}.)} \\
\end{align*}
Thus, $\plan$ is an optimal plan because $\plan^\ast$ is optimal.
\end{proof}
Therefore, from now on, we assume that all numbers in any input instance of \mutliplicationOrdering are greater than or equal to 2.

\section{Bounds on cost}
\label{section:bounds}

In this section, we introduce new lower and upper bounds on the cost of plans on a fixed instance which, in the following, will allow us to prove approximation ratios for approximation algorithms.
First, we introduce the \textsc{Pacman}-plan (denoted $\pacplan$) which has the highest cost for any instance.
Specifically, given an instance $\instance=\{\s_1,\dots,\s_n\}$ on $n$ elements such that $\s_1\geq \cdots\geq \s_n$, the first multiplication is the product of $s_1$ and $s_2$ and for $i\geq 2$,
the $i^{th}$ multiplication is the product of $s_{i+1}$ with the product of the first $i$ elements.  At each step, the \textsc{Pacman} plan introduces a new largest number; its shape is a caterpiller. Said differently, the \textsc{Pacman}-plan for $\instance$ is $((((s_1 \times s_2) \times s_3) \times s_4) \times \cdots ) \times s_n)$. Visually:
\newcommand{\tikzinnernode}{node {$\times$}}
\begin{center}
\begin{tikzpicture}[
    level distance=5mm,
    level 1/.style={sibling distance=1cm},
    edge from parent/.style={draw, thick},
    every node/.style={inner sep=0.5mm, font=\small}
]

\node {$\times$}
    child[dotted] {
        \tikzinnernode
        child[solid] { \tikzinnernode
            child { \tikzinnernode
                    child {node {$s_1$}}
                    child {node {$s_2$}}
                }
            child {node {$s_3$}}
        }
        child[solid] { node {$s_4$} }
    }
    child { node {$s_n$}};
\end{tikzpicture}
\end{center}

\begin{example}
The \textsc{Pacman}-plan for $\set{8, 5, 3, 3}$ is $(((8 \times 5) \times 3) \times 3)$.
\end{example}

The cost of a \textsc{Pacman}-plan is as follows.
\begin{proposition}\label{prop:eval_pacman}
Let $\instance = \{\s_i\}^n_{i=1}$ be an instance of \mutliplicationOrdering such that $\s_1\geq \cdots\geq \s_n$ and let $\pacplan$ be the \textsc{Pacman}-plan on $\instance$. Then, by definition, $$\costoutwithroot{\pacplan} = \uao{i=2}{n}{\sum} \left(\uao{j=1}{i}{\prod} \s_j \right) + \uao{i=1}{n}{\sum} \s_i = 
\uao{i=1}{n}{\sum} \left(\s_i + \uao{j=1}{i}{\prod} \s_j \right) -\s_1.$$
\end{proposition}

\begin{proposition}
\label{prop:worst strategy}
The \textsc{Pacman}-plan has the highest achievable cost.
\end{proposition}

\begin{proof}
For an instance $\instance$ on $n$ elements, we compare the \textsc{Pacman}-plan on $\instance$, denoted $\pacplan$, to another (arbitrary) plan $\plan$. Both plans count the leaves and the root (using $\costoutwithrootEmpty$), so we will only compare the cost of the multiplications at the $n-2$ internal nodes (i.e., non-root and non-leaf nodes).

We will argue that there is a bijection between the internal nodes of $\pacplan$ and the internal nodes of $\plan$ such that for all internal nodes of $\pacplan$, the evaluation of this node is at least as large as the evaluation of the associated node of $\plan$. 
To do this, for each plan, we order the multiplications in that plan so that a multiplication at a parent node comes after the multiplications corresponding to any of its children (and hence descendants). We call such an order a {\em hierarchy-respecting} order. We use the following claim.
\begin{claim}\label{clm:glue}
In any hierarchy-respecting order, the $i^{th}$ multiplication corresponds to the product of at most $i+1$ elements.
\end{claim}
\begin{cproof}
Suppose each time we multiply two numbers together we use a piece of glue to connect them. For example, a product of two elements contains one piece of glue and a product of four elements contains three pieces of glue. Since we use exactly one new piece of glue for each multiplication, at the $i^{th}$ step we can produce a product corresponding to a number using at most $i$ pieces of glue. Hence, it can be the product of at most $i+1$ elements, which proves the claim.
\end{cproof}

Observe that the $i^{th}$ multiplication in the (unique) hierarchy-respecting order of $\pacplan$ is the product of the $i+1$ largest elements. For a hierarchy-respecting order for $\plan$, by Claim \ref{clm:glue}, the $i^{th}$ multiplication can be the product of at most $i+1$ elements, which is clearly at most the product of the largest $i+1$ elements corresponding to the $i^{th}$ multiplication in $\pacplan$. 
\end{proof}

\begin{lemma}
\label{lemma:preciseBoundOnWorstPlan}
Let $\instance$ be an instance of \mutliplicationOrdering such that all numbers are at least $2$ and let $\pacplan$ be a \textsc{Pacman}-plan on $\instance$. Then $\costout{\pacplan} \leq \frac{1}{\min(\instance)-1} \underset{\s\in\instance}{\prod}\s + \underset{\s\in \instance}{\sum} \s - \max(\instance)$.
\end{lemma}
\begin{proof}
Wlog, assume $\instance = \{\s_i\}^n_{i=1}$ and $\s_1\geq\cdots\geq \s_n$ and let $P= \underset{\s\in\instance}{\prod} \s$. 
By Proposition \ref{prop:eval_pacman} we have
\begin{align*}
    \costoutwithroot{\pacplan} 
        =\; \uao{i=1}{n}{\sum} \s_i + \uao{i=1}{n}{\sum} \uao{j=1}{i}{\prod} \s_j -\s_1.
\end{align*}
Note that since all $\s\in\instance$ respect $\s\geq \s_n$, then $\forall i\in[n]$, $\uao{j=i+1}{n}{\prod} \s_j \geq \s_n^{n-i}$.  Therefore,
 \begin{align*}
    \uao{i=1}{n}{\sum} \uao{j=1}{i}{\prod} \s_j
        =\; \uao{i=1}{n}{\sum} \frac{\uao{j=1}{n}{\prod} \s_j}{\uao{j=i+1}{n}{\prod} \s_j}
        \leq\; P \uao{i=1}{n}{\sum} \frac{1}{\s_n^{n-i}}
        =\; P \uao{i=0}{n-1}{\sum} \frac{1}{\s_n^i}
        \leq\; P \uao{i=0}{+\infty}{\sum} \frac{1}{\s_n^i}
        =\; \frac{\s_n}{\s_n-1} P.
\end{align*}
Last inequality holds since $\s_n\geq 2$ and those the sum is converging.
Therefore, $\costout{\pacplan} = \costoutwithroot{\pacplan} - \eval{\pacplan} \leq \uao{i=1}{n}{\sum} \s_i-\s_1 + \frac{\s_n}{\s_n-1}P -P \leq \frac{1}{\min(\instance)-1}\underset{\s\in\instance}{\prod} \s + \uao{i=1}{n}{\sum} \s_i -\max(\instance)$.
\end{proof}

\begin{corollary}
\label{cor:boundOnWorstPlan}
Let $\instance$ be an instance of \mutliplicationOrdering such that all numbers are at least $2$ and let $\pacplan$ be a \textsc{Pacman}-plan on $\instance$. Then $\costout{\pacplan} \leq \underset{\s\in\instance}{\prod}\s + \underset{\s\in \instance}{\sum} \s - \max(\instance)$.
\end{corollary}

\begin{lemma}\label{lem:SumAtMostProduct}
Let $\s_1,\dots,\s_n$ be some integers greater or equal to $2$, then $\underset{i=1}{\overset{n}{\sum}} \s_i \leq \underset{i=1}{\overset{n}{\prod}} \s_i$.
\end{lemma}

\begin{proof}
For $n=2$, 
\begin{align*}
    \s_1\s_2 = (\s_1-1)\s_2 +\s_2 = (\s_1-1)(\s_2-1) + \s_2 +\s_1 -1 \geq \s_2+\s_1
\end{align*}
Thus, the property is true for $n=2$. Assume the property is true for $n$ and let $\s_1,\dots,\s_{n+1}$ be $n+1$ integers greater than $2$. Then 
\begin{align*}
    \underset{i=1}{\overset{n+1}{\sum}} \s_i = \s_{n+1} + \underset{i=1}{\overset{n}{\sum}} \s_i
         \leq \s_{n+1} + \underset{i=1}{\overset{n}{\prod}} \s_i
         \leq (\s_{n+1})\left(\underset{i=1}{\overset{n}{\prod}} \s_i\right)
         \leq \underset{i=1}{\overset{n+1}{\prod}} \s_i.
\end{align*}
The first inequality holds by the induction hypothesis and the second one holds by the case for $n=2$. Then the property holds.
\end{proof}

\begin{proposition}
\label{proposition:boundby2pi}
Consider an instance $S$ of \mutliplicationOrdering such that all numbers are at least $2$, then for all plans $\plan$ on $\instance$, $\costout{\plan}\leq 2\eval{\plan}$.
\end{proposition}
\begin{proof}
Let $\pacplan$ be a \textsc{Pacman}-plan on $\instance$. By \Cref{prop:worst strategy}, $\costout{\plan} \leq \costout{\pacplan}$ and using \Cref{cor:boundOnWorstPlan} and \Cref{lem:SumAtMostProduct} we have $\costout{\pacplan}\leq \underset{\s\in\instance}{\prod} \s + \underset{\s\in\instance}{\prod} \s$. So $\costout{\plan} \leq 2\eval{\plan}$.
\end{proof}

\begin{lemma}
\label{lemma:lowerBoundOfOPT}
Let $\instance$ be an instance of \mutliplicationOrdering on at least three elements and let $\plan=\plan_1 \times\plan_2$ be a plan on $\instance$. Then $\costout{\plan}\geq \eval{\plan_1} + \eval{\plan_2} + \underset{\s\in \instance}{\sum} \s - \s_{\max}$.
\end{lemma}
\begin{proof}
Assume that both $\plan_1$ and $\plan_2$ are plans on at least two elements. Then no leaves are accounted in $\eval{\plan_1}$ nor in $\eval{\plan_2}$. Therefore 
$\costout{\plan} \geq \underset{\s\in \instance}{\sum} \s + \eval{\plan_1} + \eval{\plan_2} \geq \underset{\s\in \instance}{\sum} \s + \eval{\plan_1} + \eval{\plan_2} - \s_{\max}$.\\
Otherwise, wlog, assume $\plan_1$ is a plan on only one element $\s_0$ thus $\plan_2$ is a plan on at least two elements. Then
$\costout{\plan} = \costout{\plan_2} + \eval{\plan_1} + \eval{\plan_2}
            \geq \underset{\s\in \instance\backslash\{\s_0\}}{\sum} \s + \eval{\plan_1} + \eval{\plan_2}
            \geq \eval{\plan_1} + \eval{\plan_2} + \underset{\s\in \instance}{\sum} \s -\s_{\max}$.
\end{proof}

    \subsection{{\normalfont{\textsc{Half}}}-plan}
    
In this subsection, we introduce the naive notion of a \textsc{Half}-plan (denoted $\halfplan$) and prove an upper bound on its cost. These results are used to prove approximation results such as an FPTAS for the cost with the root in \Cref{thm:FPTAS with root for k>=2}. A \textsc{Half}-plan on an instance $\instance$ is a plan that puts half of the input numbers on one side and the rest of the numbers on the other side. Therefore if the \textsc{Half}-plan on instance $\instance$ on $n$ elements, $\halfplan=\plan_1\times\plan_2$, follows $X$, then both $X$ and $\instance\backslash X$ have at least $\lfloor\frac{n}{2}\rfloor$ elements. Notice that an instance can have several {\textsc{Half}}-plans.
\begin{example}
An example of a {\textsc{Half}}-plan for $\{2, 2, 3, 3, 5, 8, 9\}$ is $((2 \times 8) \times 3) \times (5\times (3\times (2 \times 9))$.
\end{example}

\begin{lemma}\label{lemma:costOfHalfPlan}
Let $\instance$ be an instance of \mutliplicationOrdering, then
$\costout{\halfplan} \leq \frac{6}{\min(\instance)^{\frac{n}{2}-1}} \underset{\s\in\instance}{\prod} \s$.
\end{lemma}
\begin{proof}
Let $\halfplan = \plan_1 \times \plan_2$ be an \textsc{Half}-plan on $\instance$.
Firstly, notice that since all numbers in $\plan_1$ are greater than or equal to $\min(\instance)$ and since $\plan_1$ contains at least $\frac{n}{2}-1$ numbers, we have, $\eval{\plan_1}\geq \min(\instance)^{\frac{n}{2}-1}$.
Since $\underset{\s\in\instance}{\prod} \s =  \eval{\halfplan}=\eval{\plan_1}\cdot\eval{\plan_2}$, we have,
\begin{align*}
    \eval{\plan_2} = \frac{\eval{\halfplan}}{\eval{\plan_1}}
        \leq \frac{\underset{\s\in\instance}{\prod} \s}{\min(\instance)^{\frac{n}{2}-1}}.
\end{align*}

Similarly, $\eval{\plan_1} \leq \frac{\underset{\s\in\instance}{\prod} \s}{\min(\instance)^{\frac{n}{2}-1}}$. Therefore,
\begin{align*}
    \costout{\plan} =\; \costout{\plan_1} + \costout{\plan_2} + \eval{\plan_1} + \eval{\plan_2}
        \leq\; 3(\eval{\plan_1} + \eval{\plan_2})
        \leq\; \frac{6}{\min(\instance)^{\frac{n}{2}-1}}\underset{\s\in\instance}{\prod} \s.
\end{align*}
First inequality uses \Cref{proposition:boundby2pi} on $\plan_1$ and $\plan_2$ and the second inequality uses the above inequalities.~\end{proof}

\begin{lemma}
\label{lemma:halfplaExponentialyBetterThanPacman}
For any instance $\instance$ of \mutliplicationOrdering, $\costout{\halfplan} \leq \frac{24}{2^\frac{n}{2}}\costout{\pacplan}$.
\end{lemma}
\begin{proof}
Let $\instance = \{\s_i\}^n_{i=1}$ be an instance of \mutliplicationOrdering such that $\s_1\geq \cdots\geq \s_n \geq 2$ then, by \Cref{prop:eval_pacman}, we have 
\begin{align*}
    \costout{\pacplan} = \uao{i=1}{n}{\sum} \left(\s_i + \uao{j=1}{i}{\prod} \s_j \right) -\s_1 - \underset{\s\in\instance}{\prod} \s
        \geq \uao{i=1}{n-1}{\sum} \uao{j=1}{i}{\prod} \s_j
        \geq \uao{j=1}{n-1}{\prod} \s_j
        \geq \frac{1}{\s_n}\underset{\s\in\instance}{\prod} \s.
\end{align*}
Using \Cref{lemma:costOfHalfPlan} we have,
\begin{align*}
    \costout{\halfplan} \leq \frac{6}{\min(\instance)^{\frac{n}{2}-1}} \underset{\s\in\instance}{\prod} \s
        \leq \frac{6}{\min(\instance)^{\frac{n}{2}-2}}\costout{\pacplan}
        \leq \frac{24}{2^{\frac{n}{2}}}\costout{\pacplan}.
\end{align*}
\end{proof}

    \subsection{Trivial approximation factors}
Thanks to the bound on the cost of a plan obtained in Proposition \ref{proposition:boundby2pi}, we are able to find approximation factors for any plan. Namely, we show that no matter the chosen plan, we can prove a guarantee on its cost.

\begin{lemma}
Let $\instance$ be an instance with all input elements $\geq 2$.  Then every plan $\plan$ on $\instance$ is a $3$-approximation of \mutliplicationOrdering for the $\costoutwithrootEmpty$ objective function.
\end{lemma}
\begin{proof}
Let $\plan$ be a plan and let $\plan^\ast$ be an optimal strategy on $\instance$. Then by \Cref{proposition:boundby2pi} we have
\begin{align*}
    \costoutwithroot{\plan} \leq 3\eval{\plan} = 3\eval{\plan^\ast} \leq 3 \costoutwithroot{\plan^\ast} \leq 3\optwithroot{\instance}.
\end{align*}
\end{proof}

\begin{lemma}
Let $\instance$ be an instance of \mutliplicationOrdering such that all numbers are at least $2$ and let $P = \underset{\s\in \instance}{\prod} \s$. Then every plan on $\instance$ is a $\left(2\sqrt{P}\right)$-approximation for $\costoutEmpty$.
\end{lemma}
\begin{proof}
Let $\plan=\plan_1\times\plan_2$ be a plan on $\instance$. Since $P = \eval{\plan} = \eval{\plan_1}\cdot \eval{\plan_2}$ then either $\eval{\plan_1}$ or $\eval{\plan_2}$ is bigger than $\sqrt{P}$. Thus $\costout{\plan} = \costout{\plan_1} + \costout{\plan_2} + \eval{\plan_1} + \eval{\plan_2}\geq \sqrt{P}$. Thus $\opt{\instance}\geq \sqrt{P}$. Moreover, by property \ref{proposition:boundby2pi} since all $\s_i$'s are at least $2$, we have $\costout{\plan} \leq 2\eval{\plan} \leq 2P$. Thus $\frac{\costout{\plan}}{\opt{\instance}} \leq \frac{2P}{\sqrt{P}} \leq 2\sqrt{P}$.
\end{proof}

\section{FPTAS}
\label{section:fptas}
In this section, we present a fully polynomial-time approximation scheme (FPTAS) for both the $\costoutwithrootEmpty$ and $\costoutEmpty$ objective functions.  Given a parameter $\varepsilon>0$, the algorithm produces a $(1+\varepsilon)$ approximate solution in time polynomial in both $\frac{1}{\varepsilon}$ and in the size of the instance.

\subsection{Cost that includes the root}
First, we tackle the case of the cost that includes the final product at the root, which is easier and will be used in the FPTAS for the case in which the cost excludes the root.
\begin{theorem}
\label{thm:FPTAS with root for k>=2}
\mutliplicationOrdering for $\costoutwithrootEmpty$ has an FPTAS.
\end{theorem}
\begin{proof}
Let $\varepsilon > 0$ be the parameter in the desired approximation bound and let $\instance$ be an instance of \mutliplicationOrdering on $n$ elements. Thanks to Proposition \ref{prop:someKEquaOne}, we can apply a polynomial times preprocessing on the instance and assume that all input elements in $\instance$ are at least $2$. Then, we consider the following algorithm:
\begin{enumerate}
    \item If $\varepsilon < \frac{12}{2^{\frac{n}{2}}}$, we solve it using the exact algorithm of \Cref{thm:exact solving in 3^n}.
    \item Otherwise construct a \textsc{Half}-plan $\halfplan$ on $\instance$ (defined at the end of \Cref{section:bounds}).
\end{enumerate}

\paragraph{The algorithm outputs a $(1+\epsilon$)-approximation.}
If $\varepsilon < \frac{12}{2^{\frac{n}{2}}}$, we solve exactly the problem, the output strategy is optimal and thus is a $(1+\varepsilon)$-approximation.
Otherwise, $\varepsilon \geq \frac{12}{2^{\frac{n}{2}}}$. \Cref{lemma:costOfHalfPlan} give us that $\costoutwithroot{\halfplan} \leq \frac{6}{2^{\frac{n}{2}-1}}\eval{\halfplan} + \eval{\halfplan}$ since the smaller number in $\instance$ is at least $2$. Therefore,
\begin{align*}
    \costoutwithroot{\halfplan} \leq \; \left(\frac{12}{2^{\frac{n}{2}}} + 1\right)\eval{\halfplan}
        \leq \; (1+\varepsilon)\eval{\halfplan} 
        \leq \; (1+\varepsilon) \optwithroot{\instance}.
\end{align*}
Since $\optwithroot{\instance}\geq\underset{\s\in \instance}{\prod} \s = \eval{\halfplan}$. Hence $\halfplan$ is a $(1+\varepsilon)$-approximation of \mutliplicationOrdering using $\costoutwithrootEmpty$.

\paragraph{Complexity analysis.}
If $\varepsilon < \frac{12}{2^{\frac{n}{2}}}$ then we consider the dynamic programming of Theorem \ref{thm:exact solving in 3^n} running in time $\mathcal{O}(3^n)$. We have $n<\frac{2\log(\frac{12}{\varepsilon})}{\log(2)} \leq 2\frac{\log 3}{\log 2}\log_3(\frac{12}{\varepsilon})$. Then $3^n\leq \left(3^{\log_3(\frac{12}{\varepsilon})}\right)^{\frac{2\log 3}{\log 2}} = \left(\frac{12}{\varepsilon}\right)^{\frac{2\log 3}{\log 2}}$ which is polynomial in $\frac{1}{\varepsilon}$ and does not depend on $n$.
Otherwise, notice that constructing $\plan_1$ and $\plan_2$ is polynomial in $n$ and in $\frac{1}{\varepsilon}$ since we just split the $\s_i$'s in half and take arbitrarily ordering.
Thus, this shows that the proposed algorithm is a FPTAS algorithm.
\end{proof}

\subsection{Cost that excludes the root}
We now tackle the problem where we use the $\costoutEmpty$ objective function, which does not count the root. For this purpose, we introduce the known \ppb problem which is an \NP-complete problem first introduced in \cite{DBLP:journals/eor/NgBCK10}. \ppb takes as input a multiset of positive integers $\instance=\{\s_i\}_{i=1}^n$ and outputs a non-empty subset $X\subset \instance$ minimizing the majority of $X$ in $\instance$ denoted $\fnmax{\instance}(X) = \max\left( \underset{\s\in X}{\prod} \s,\; \underset{\s\in\instance\backslash X}{\prod} \s \right)$. \ppb has been proved having an FPTAS algorithm in section 4 of \cite{DBLP:journals/eor/NgBCK10}. We will use this result to design an FPTAS algorithm for \mutliplicationOrdering for $\costoutEmpty$.

\begin{proposition}
\label{proposition:approxproductpartition}
[See Section 4 of \cite{DBLP:journals/eor/NgBCK10}.]
There is an algorithm running in time $\mathcal{O}(n^3\log(\s_{max}) \frac{1}{\varepsilon})$ that computes a $(1+\varepsilon)$-approximation of \ppb. 
\end{proposition}

In order to make the connection between \ppb and \mutliplicationOrdering, we have to compare the majority and the weight.  We will firstly show that the \ppb is equivalent to finding a subset $X\subseteq\instance$ minimizing the weight of $X$ in $\instance$, denoted $\weight{\instance}(X) := \underset{\s\in X}{\prod} \s + \underset{\s\in\instance\backslash X}{\prod} \s$. Notice that for every plan $\plan=\plan_1\times \plan_2$ following a set $X$, $\costout{\plan} = \costout{\plan_1} + \costout{\plan_2} + \weight{\instance}(X)$.

\begin{lemma}
\label{lemma:minimizing ppo also minimize the sum}
Let $\instance$ be an instance of \ppb, if $X\subseteq\instance$ minimizes $\fnmax{\instance}(X)$ then $X$ also minimizes $\weight{\instance}(X)$.
\end{lemma}
\begin{proof}
Let $P=\underset{\s\in\instance}{\prod}\s$. Notice first that $0<\sqrt{P}\leq \fnmax{\instance}(X) \leq P$ and that $\weight{\instance}(X) = \fnmax{\instance}(X) + \frac{P}{\fnmax{\instance}(X)}$. Let $h:\begin{cases}
    [\sqrt{P},P] \rightarrow \mathbb{R}\\
    x \mapsto x + \frac{P}{x}
\end{cases}$.
Notice that $\forall x \in [\sqrt{P},P]$, $h'(x) = 1-\frac{P}{x^2} \geq 0$ so $h$ is increasing. Thus, $\forall X,Y\subseteq\instance, \fnmax{\instance}(X)\leq \fnmax{\instance}(Y) \Leftrightarrow \weight{\instance}(X)\leq \weight{\instance}(Y)$. Therefore, if $X$ minimizes $\fnmax{\instance}(X)$, $X$ also minimizes $\weight{\instance}(X)$.
\end{proof}

\begin{lemma}
\label{lemma:approximate ppo also approximate the sum}
Let $\instance$ be an instance of \ppb, $\varepsilon>0$, $X\subseteq\instance$ minimizing $\fnmax{\instance}$, and let $Y\subseteq\instance$ be a $(1+\varepsilon)$-approximation of $\fnmax{\instance}$. Then $Y$ is a $(1+\varepsilon)$-approximation of $\weight{\instance}$.
\end{lemma}
\begin{proof}
Firstly, since $Y$ is a $(1+\varepsilon)$-approximation of $\fnmax{\instance}$, $\fnmax{\instance}(Y)\leq (1+\varepsilon)\fnmax{\instance}(X)$.
Secondly, notice that $0<\fnmax{\instance}(X)\leq \fnmax{\instance}(Y)\leq (1+\varepsilon)\fnmax{\instance}(X)$. Thus $t=\frac{\fnmax{\instance}(Y)}{\fnmax{\instance}(X)}\in [1,1+\varepsilon]$. Let $P=\underset{\s\in \instance}{\prod} \s$ and $\alpha = \frac{P}{\fnmax{\instance}(X)^2}$, thus
\begin{align*}
    \frac{\weight{\instance}(Y)}{\weight{\instance}(X)}
    = \frac{t\cdot \fnmax{\instance}(X) + \frac{P}{t\cdot \fnmax{\instance}(X)}}{\fnmax{\instance}(X) + \frac{P}{\fnmax{\instance}(X)}}
    = \frac{t+\frac{\alpha}{t}}{1+\alpha}
    \leq \frac{1+\varepsilon+\alpha}{1+\alpha}\leq 1+\varepsilon.
\end{align*}
Since $X$ is minimizing $\fnmax{\instance}$, by Lemma \ref{lemma:minimizing ppo also minimize the sum}, $X$ also minimizing $\weight{\instance}$. This concludes the proof. 
\end{proof}

\begin{corollary}
\label{corollary:approxsumproductpartition}
There is an algorithm running in time $\mathcal{O}(n^3\log(\s_{max}) \frac{1}{\varepsilon})$ that computes a $(1+\varepsilon)$-approximation of the minimization of $\weight{\instance}$ on an instance $\instance$ of \ppb. 
\end{corollary}
\begin{proof}
By \Cref{proposition:approxproductpartition} and \Cref{lemma:approximate ppo also approximate the sum}.
\end{proof}

\begin{theorem}
\label{thm:FPTASwithoutRoot}
\mutliplicationOrdering for $\costoutEmpty$ has an FPTAS.
\end{theorem}
\begin{proof}
Let $\instance$ an instance of \mutliplicationOrdering and $\varepsilon>0$ a fixed constant and let $\varepsilon' = \sqrt{1+\varepsilon}-1$.  As in the FPTAS presented previously, we first perform a preprocessing step so we can assume all numbers in $\instance$ are greater or equal to $2$. We consider the following algorithm:
\begin{enumerate}
    \item Compute a set $X\subseteq\instance$ that is a $(1+\varepsilon')$-approximation of $\weight{\instance}$.
    \item Return $\plan = \plan_1\times\plan_2$ where $\plan_1$ (resp. $\plan_2$) is computed via the FPTAS for $\costoutwithrootEmpty$ with $\varepsilon'$ on instance $X$ (resp. $\instance\backslash X$).
\end{enumerate}

\paragraph{This algorithm outputs a $(1+\varepsilon)$-approximation}
Firstly, using the bound found in the proof of \Cref{thm:FPTAS with root for k>=2}, we have that $\costoutwithroot{\plan_1} \leq (1+\varepsilon')\eval{\plan_1}$ and the same holds for $\plan_2$. Thus 
\begin{align*}
    \costout{\plan} = \costoutwithroot{\plan_1} + \costoutwithroot{\plan_2} \leq (1+\varepsilon')\left(\eval{\plan_1} + \eval{\plan_2}\right)
    = (1+\varepsilon')\weight{\instance}(X).
\end{align*}
Let $\plan^{opt} = \plan^{opt}_1\times \plan^{opt}_2$ be an optimal plan on $\instance$, $X^{opt}$ be the set followed by $\plan^{opt}$ and $X^\ast\subseteq\instance$ be the subset minimizing $\weight{\instance}$. Thus $\weight{\instance}(X)\leq (1+\varepsilon')\weight{\instance}(X^\ast)$ using  Lemma \ref{lemma:approximate ppo also approximate the sum} and $\weight{\instance}(X^\ast) \leq \weight{\instance}(X^{opt})$ since $X^\ast$ is optimal. Therefore,
\begin{align*}
    \costout{\plan} \leq (1+\varepsilon')\weight{\instance}(X)
    \leq (1+\varepsilon')^2\weight{\instance}(X^{opt})
    \leq (1+\varepsilon)\costout{\plan^{opt}}
    = (1+\varepsilon)\opt{\instance}.
\end{align*}
This concludes the proof that the algorithm is a $(1+\varepsilon)$-approximation of \mutliplicationOrdering for $\costoutEmpty$.

\paragraph{Complexity}
Step 1 can be implemented in polynomial time in $n$, $\frac 1 \epsilon$ and the number of bits in the $\s_i$'s thanks to \Cref{proposition:approxproductpartition}. This is polynomial in the size of the input and in $\frac{1}{\varepsilon}$ since $\s_i$'s are given in binary, thus the $\log$ of the biggest $\s_i$ is its size. 
\end{proof}

\section{Link to {\normalfont \ppb}} \label{section:linkToProdPart}
The purpose of this section is to highlight links between \mutliplicationOrdering and \ppb motivated by the FPTAS for the $\costoutEmpty$ objective described in Theorem \ref{thm:FPTASwithoutRoot}. More precisely, we prove that to reach optimality for \mutliplicationOrdering, the first partition from the top of the plan has to be a good approximation of \ppb which gives rise to conditions to achieve optimality.  However, \mutliplicationOrdering and \ppb are not equivalent in the general case (but are when all input elements have the same value). Since instances of size two are obvious (no choices to make), we can assume that all considered instances in this section have at least three input elements.

\begin{lemma}
\label{lemma:product partition is 2-APX}
Let $\instance$ be an instance of \mutliplicationOrdering and let $X\subset\instance$ be an $\gamma\geq 1$ approximation of \ppb on $\instance$. Then every plan following $X$ is a $\left(\gamma\frac{\min(\instance)}{\min(\instance) -1}\right)$-approximation for $\costoutEmpty$.

In particular, since each elements are greater or equal to $2$, the plan is always a $(2\gamma)$-approximation.
\end{lemma}
\begin{proof}
We first introduce three plans. Let $\plan=\plan_1\times\plan_2$ be a plan following $X$ and $\plan^\ast=\plan^\ast_1\times\plan^\ast_2$ be a plan following $X^\ast$ with $X^\ast$ being an optimal solution of \ppb on $\instance$ and let $\plan^{opt}=\plan^{opt}_1\times\plan^{opt}_2$ be an optimal solution of \mutliplicationOrdering on $\instance$.
Then, let $a=\min(X)$ and $b=\min(\instance\backslash X)$. Wlog we can assume $a = \min(\instance)$. Thus, $a\leq b$ and then $\frac{1}{a-1}\geq \frac{1}{b-1}$. Therefore, using \Cref{lemma:preciseBoundOnWorstPlan} on $\plan_1$, we have
\begin{align*}
    \costout{\plan_1}
        \leq \frac{1}{a-1}\eval{\plan_1} + \underset{\s\in X}{\sum}\s - \max(X)
\end{align*}
and using \Cref{lemma:preciseBoundOnWorstPlan} on $\plan_2$, we have
\begin{align*}
    \costout{\plan_2}
        \leq \frac{1}{b-1}\eval{\plan_2} + \underset{\s\in \instance\backslash X}{\sum}\s - \max(\instance\backslash X)
        \leq \frac{1}{a-1}\eval{\plan_2} + \underset{\s\in \instance\backslash X}{\sum}\s - \max(\instance\backslash X).
\end{align*}

Then,
\begin{align*}
    \costout{\plan} =&\; \eval{\plan_1} + \eval{\plan_2} + \costout{\plan_1} + \costout{\plan_2}\\
        \leq&\;  \left(1+\frac{1}{a-1}\right)(\eval{\plan_1} + \eval{\plan_2}) + \underset{\s\in X}{\sum} \s -\max(X) + \underset{\s\in\instance\backslash X}{\sum} \s - \max(\instance\backslash X)\\
        \leq&\; \gamma\frac{a}{a-1}(\eval{\plan^\ast_1} + \eval{\plan^\ast_2}) + \underset{\s\in \instance}{\sum} \s - \max(\instance)\\
        \leq&\; \frac{\gamma\cdot a}{a-1}\left(\eval{\plan^\ast_1} + \eval{\plan^\ast_2} + \underset{\s\in \instance}{\sum} \s - \max(\instance)\right)\\
        \leq&\; \frac{\gamma\cdot a}{a-1}\left(\eval{\plan^{opt}_1} + \eval{\plan^{opt}_2} + \underset{\s\in \instance}{\sum} \s - \max(\instance)\right)\\
        \leq&\; \frac{\gamma\cdot a}{a-1}\cdot\costout{\plan^{opt}}\\
        =&\; \frac{\gamma\cdot \min(\instance)}{\min(\instance)-1}\cdot\opt{\instance}.
\end{align*}
The first inequality uses above remarks. The second one is because $\max(\instance)$ is either equal to $\max(X)$ or to $\max(\instance\backslash X)$ and by using the assumption that $X$ is a $\gamma$-approximation of \ppb on $\instance$. The third one is by observing that $\frac{\gamma\cdot a}{a-1}\geq 1$ and $\underset{\s\in \instance}{\sum} \s - \max(\instance)\geq 0$. The fourth one is because $X^\ast$ is an optimal solution of \ppb and the fifth one is using \Cref{lemma:lowerBoundOfOPT}.
\end{proof}

\begin{corollary}
\label{cor:producPartitionPlanTendToOPT}
Let $\instance$ be an instance of \mutliplicationOrdering and let $\plan$ be a plan following an optimal solution to \ppb on \instance. Then the cost of $\plan$ tend to the optimal cost when $\min(\instance)$ tends to infinity.
\end{corollary}
\begin{proof}
Let $a = \min(\instance)$ and using \Cref{lemma:product partition is 2-APX} with $\gamma = 1$, we have
\begin{align*}
    \costout{\plan}
    \leq \frac{a}{a -1} \opt{\instance}
    \underset{a\rightarrow +\infty}{\longrightarrow} \opt{\instance}.
\end{align*}
\end{proof}

\begin{lemma}
\label{lemma:notBeing2approwOfPPImpliesNotBeingOPT}
Let $\instance$ be an instance of \mutliplicationOrdering and let $X$ be a subset of $\instance$ being worse than or equal to a $\left(\frac{\min(\instance)}{\min(\instance)-1}\right)$-approximation of \ppb. Then no plan following $X$ is optimal.
\end{lemma}
\begin{proof}
Let $X^\ast$ be a subset of $\instance$ being optimal for \ppb and let $\plan^\ast=\plan^\ast_1\times\plan^\ast_2$ following $X^\ast$. Let $\plan=\plan_1\times\plan_2$ following $X$. Then,
\begin{align*}
    \costout{\plan} \geq&\; \eval{\plan_1} + \eval{\plan_2} + \underset{\s\in \instance}{\sum}\s -\max(\instance)\\
            \geq&\; \frac{\min(\instance)}{\min(\instance)-1} \left( \eval{\plan^\ast_1} + \eval{\plan^\ast_2} \right) + \underset{\s\in \instance}{\sum}\s -\max(\instance)\\
            >&\; \frac{\min(\instance)}{\min(\instance)-1} \left( \eval{\plan^\ast_1} + \eval{\plan^\ast_2} \right) + \underset{\s\in X^\ast}{\sum} \s -\max(X^\ast) + \underset{\s\in \overline{X^\ast}}{\sum} \s - \max(\overline{X^\ast})\\
            \geq&\; \eval{\plan^\ast_1} + \eval{\plan^\ast_2} + \costout{\plan^\ast_1} + \costout{\plan^\ast_2}\\
            =&\; \costout{\plan^\ast}\\
            =&\; \opt{\instance}.
\end{align*}
First inequality uses \Cref{lemma:lowerBoundOfOPT}, second uses the fact that $X$ is worst or equal to a $\left(\frac{\min(\instance)}{\min(\instance)-1}\right)$-approximation of \ppb. The third one uses the assumption that all elements are greater or equal to $2$. The fourth inequality uses \Cref{lemma:preciseBoundOnWorstPlan} on $\plan^\ast_1$ and on $\plan^\ast_2$.
\end{proof}

\Cref{lemma:product partition is 2-APX}, \Cref{lemma:notBeing2approwOfPPImpliesNotBeingOPT}, \Cref{prop:allkEqualInnlogn} and the FPTAS algorithm in \Cref{thm:FPTASwithoutRoot} highlighted strong connections between \mutliplicationOrdering and \ppb. Therefore, we introduce the notion of \ppb-plan, which is a plan following the strategy consisting of spitting the instance using an optimal solution to \ppb at each step. A direct consequence of Lemma \ref{lemma:product partition is 2-APX} is that every \ppb-plan are $2$-approximation of \mutliplicationOrdering. However, the following proposition contradict optimality of \ppb-plans.
\begin{proposition}
\label{prop:productPartitionCounterExample}
\ppb-plans are not always optimal.
\end{proposition}
\begin{proof}
We highlight a counter example following this logic: Let $t\geq 2$, $c\geq 2$ and $\instance = \{ b,\dots,b, b^t-1, b^t-1,b^t+2\}$, where there are $3t$ occurrences of $b$ in the instance. So, for $t=3$ and $b=2$, $\instance=\{2,2,2,2,2,2,2,2,2,7,7,10\}$. Let $\plan$ be the plan following the described strategy. One can check that there is only one optimal solution of \ppb on $\instance$ which is $X=\{b^t-1, b^t-1, b^t+2\}$ and $\instance\backslash X$ is the set containing all the $b$'s. Let $X^\ast = \{b,\dots,b,b^T-1,b^t+2\}$ containing $t$ occurrences of $b$. Let $\plan^\ast$ be the plan following $X^\ast$ at first step and then following the most balanced splitting for the subplan. Let $\plan$ be the plan following the \ppb strategy, then $\plan$ follows $X$. The best valid counter example found is for $t=2$ and $b=4$. Thus $\plan^\ast$ is not following the \ppb plan since its first split is not optimal for \ppb. In fact, one can check that $\fnmax{\instance}(X) = 4096 < 4320 = \fnmax{\instance}(X^\ast)$. One can also check that $\costout{\plan} = 8603 > 8504 = \costout{\plan^\ast}$. which gives a ration of $\frac{8603}{8504}\simeq 1.0116$. Thus $\plan$ is not optimal. The values found here were verified with computers.
\end{proof}

\begin{proposition}
\label{prop:allkEqualInnlogn}
Let $\instance = \{\s_i\}^n_{i=1}$ be an instance of \mutliplicationOrdering with $s_1 = s_2 = \dots = s_n = s$.
Then we can solve \mutliplicationOrdering in $\mathcal{O}(n\log n)$ time using \ppb.
\end{proposition}
\begin{proof}
Firstly, we can assume that $\s\geq 2$ since if $\s=0$ every plan will have a cost of $0$ and if $\s=1$ the optimal plan will be to successively multiply $1$ to the smallest other number \Cref{prop:someKEquaOne}, namely any other number. Consider the following inductive algorithm:
\begin{enumerate}
    \item If $n=1$ return the unique plan on a single leaf.
    \item Otherwise, create $X$, a subset containing half of the numbers (for instance $X=\{\s_1,\dots, \s_{\lceil \frac{n}{2} \rceil}\}$) and construct a plan $\plan=\plan_1\times\plan_2$ following $X$ such that $\plan_1$ (resp. $\plan_2$) are constructed via this algorithm on the instance $X$ (resp. $\instance\backslash X$).
\end{enumerate}
We will show that if a plan $\plan$ does not split the instance in half then it cannot be optimal. Let $X^\ast$ splitting the instance in half and $X$ not doing that. Wlog, assume $|X^\ast|\geq |\instance\backslash X^\ast|$ and $|X|\geq |\instance\backslash X|$. Thus $|X|>|X^\ast|$. Therefore, $\s^{|X|} \geq \s\cdot \s^{|X^\ast|}$ and thus $\fnmax{\instance}(X) \geq \s\cdot\fnmax{\instance}(X^\ast)$ since $\underset{\s\in X}{\prod} \s = \underset{\s\in X}{\prod} \s = \s^{|X|}$. Now, note that since all $\s_i$ are equal then $X^\ast$ is the best product partition thus $X$ is worst or equal to a $k$-approximation of \ppb. Since $\s\geq 2$ and by \Cref{lemma:notBeing2approwOfPPImpliesNotBeingOPT}, every plan not splitting in half the instance is not optimal. Thus, every optimal plan split the instance in half at each step.

The complexity analysis is simple, for an instance of size $n$, let $T(n)$ be the complexity to compute the solution. By induction, $T(n) = 2T(\frac{n}{2}) + n$ since the cost of creating $X$ is linear in $n$. By the classic master theorem, $T(n) = \mathcal{O}(n\log n)$.
\end{proof}

\section{Restriction when numbers are powers of a single basis}
\label{section:restriction}

The \NP-hardness proof in \cite{DBLP:conf/pods/ScheufeleM97} uses a reduction where they construct an instance of \mutliplicationOrdering with distinct prime numbers in order to make distinctions between the solutions. In contrast, in this section, we consider the special case of instances with regularity. In particular, we consider instances where all the numbers $\s_i$ are powers of the same base $b \geq 2$. For instance, for $b = 4$, an instance could be $4^5, 4^{12}, 4^{12}, 4^{459}$. Naturally, when numbers $\s_i$ are written in binary, it corresponds to the exponents to be written in unary, which is the case we consider now.

\subsection{Exponents in unary}

\begin{proposition}
\label{prop:optimal stratgy for power of c}
Let $b\geq 4$ and let $\instance = \{\s_i\}^n_{i=1}$ be an instance of positive powers of $b$, namely, $\s_i = b^{a_i}$ with $a_i>0$. Let $X$ be a partition of the $\s_i$'s which is not optimal for \ppb. Then every plan following $X$ is not optimal.
\end{proposition}
\begin{proof}
Let $X^\ast$ be an optimal partition of \ppb and let $\plan^\ast = \plan^\ast_1\times\plan^\ast_2$ (resp. $\plan=\plan_1\times\plan_2$) be a plan following $X^\ast$ (resp. $X$). Wlog, assume $\underset{\s\in X^\ast}{\prod} \s \geq \underset{\s\in\instance\backslash X^\ast}{\prod} \s$ and $\underset{\s\in X}{\prod} \s \geq \underset{\s\in\instance\backslash X}{\prod} \s$. Because $b\geq 2$ and all $a_i\geq 1$, all $\s_i\geq 2$, therefore
\begin{align*}
    \opt{\instance} \leq&\; \costout{\plan_1^\ast} + \costout{\plan_2^\ast} + \eval{\plan_1^\ast} + \eval{\plan_2^\ast}\\
        \leq&\; 2(\eval{\plan_1^\ast} + \eval{\plan_2^\ast} ) + \underset{\s\in X^\ast}{\sum} \s - \max(X^\ast) + \underset{\s\in\overline{X^\ast}}{\sum} \s - \max(\overline{X^\ast})\\
        \leq&\; 4\underset{\s\in X^\ast}{\prod} \s + \underset{\s\in \instance}{\sum} \s - \max(\instance).
\end{align*}
Second inequality uses \Cref{cor:boundOnWorstPlan} on $\plan^\ast_1$ and $\plan^\ast_2$. Since $X$ is not optimal for \ppb but $X^\ast$ is, we have $\underset{\s\in X}{\prod} \s > \underset{\s\in X^\ast}{\prod} \s$, which implies $\underset{\s\in X}{\prod} \s \geq b\underset{\s\in X^\ast}{\prod} \s$. Thus
\begin{align*}
    \costout{\plan} \geq\; \eval{\plan_1} + \eval{\plan_2} + \underset{\s\in \instance}{\sum}\s -\max(\instance)
            >\; \underset{\s\in X}{\prod} \s + \underset{\s\in \instance}{\sum}\s -\max(\instance)
            \geq\; b \underset{\s\in X^\ast}{\prod} \s + \underset{\s\in \instance}{\sum}\s -\max(\instance).
\end{align*}
First inequality uses lemma \ref{lemma:lowerBoundOfOPT}. Then, $\costout{\plan}>\opt{\instance}$. Thus, every plan not following an optimal product partition on the instance are not optimal.
\end{proof}

\begin{theorem}
\label{theorem:power}
Let $b\geq 4$ and let $\instance=\{\s_1,\dots,\s_n\}$ be an instance of positive power of $b$, namely, $\s_i = b^{a_i}$ with $a_i>0$. Then we can solve \mutliplicationOrdering on $\instance$ in polynomial time.
\end{theorem}
\begin{proof}
Consider the following strategy: at each level, find an optimal product partition of the instance and create a plan following it. This strategy is optimal by \Cref{prop:optimal stratgy for power of c}. Let $I(X)$ be the set of index of the numbers in $X$, namely, $I(X) = \{i\in[n],\; s_i\in X\}$. Then, minimizing $\weight{\instance}(X)$ is the same as minimizing $\fnmax{\instance}$ by \Cref{lemma:minimizing ppo also minimize the sum} which is the same  as minimizing $\underset{i\in I(X)}{\prod} \s_i + \underset{i\in [n]\backslash I(X)}{\prod} \s_i$ which  which is the same as minimizing $\max\left(\underset{i\in I(X)}{\sum} a_i,\; \underset{i\in[n]\backslash I(X)}{\sum} a_i  \right)$ by taking the $\log$. Since the number are given in binary, the exponents $a_i$'s are in unary. This is exactly an instance of \partition with input written in unary which is solvable in polytime (see \cite{DBLP:conf/coco/Karp72}).
\end{proof}

\subsection{Exponents in binary}

Now we assume that the exponents are written in binary. For example, in the instance $4^5$, $4^{12}$, $4^{12}$, $4^{459}$, the exponents 5, 12, 12, 459 are written in binary. In that case, the problem becomes \NP-complete.

\begin{theorem}
\label{theorem:powersexponentsinbinary}
\mutliplicationOrdering when all numbers are power of a base $b\geq 2$ and exponents written in \emph{binary} is N\P-complete.
\end{theorem}
\begin{proof}
For the \NP-membership, the certificate is just a plan $\plan$. For the \NP-hardness proof, we reduce \partition to  \mutliplicationOrdering. \partition is a problem that takes as input a mutliset of integers $A=\{a_1,\dots,a_n\}$ and the goal is to determine whether there exists a subset $X\subset A$ such that $\underset{a\in X}{\sum}a=\underset{a\in\overline{X}}{\sum} a$, it has been proved \NP-complete in \cite{DBLP:conf/coco/Karp72}. 
Let $A =\{a_1,\dots,a_n\}$ be an instance of \partition with no zeros and let $S = \underset{a\in A}{\sum} a$. Let $\instance=\{\s_1,\dots,\s_n\}$ and $\beta = 6\cdot b^{\frac{3}{2}S}$ be an instance of \mutliplicationOrdering such that $\forall i \in \{1,\dots,n\}$, $\s_i = c^{3a_i} \geq 2$. For all $A'\subseteq A$, we denote by $I(A') = \{i\in[n],\; a_i\in A'\}$.

Let $A_1,A_2 \subseteq A$ such that $A_1 \bigsqcup A_2 = A$. Then let $\plan=\plan_1\times\plan_2$ a plan following $A_1$. Wlog assume that $\plan_1$ is a plan on $A_1$ and $\plan_2$ a plan on $A_2$. If $\underset{a\in A_1}{\sum} a = \underset{a\in A_2}{\sum} a = \frac{S}{2}$, thus we have
\begin{align*}
    \eval{\plan_1} \leq  \underset{i\in I(A_1)}{\prod} \s_i
            \leq b^{\underset{i\in I(A_1)}{\sum} 3a_i}
            \leq b^{\frac{3}{2}S}.
\end{align*}
Similarly, we have, $\eval{\plan_2} \leq c^{\frac{3}{2}S}$. Using \Cref{proposition:boundby2pi}, we have
\begin{align*}
    \costout{\plan} \leq  \costout{\plan_1} + \costout{\plan_2} + \eval{\plan_1} + \eval{\plan_2}
    \leq 3\left( \eval{\plan_1} + \eval{\plan_2}\right)
    \leq 6\cdot b^{\frac{3}{2}S}
    \leq \beta.
\end{align*}
Otherwise, $A_1$ and $A_2$ are such that $\underset{a\in A_1}{\sum} a \neq \underset{a\in A_2}{\sum} a$. Wlog, we can assume that $\underset{a\in A_1}{\sum} a < \underset{a\in A_2}{\sum} a$. So $\underset{a\in A_2}{\sum} a \geq \frac{S}{2}+1$. Thus,
\begin{align*}
    \costout{\plan} 
    \geq \eval{\plan_1} + \eval{\plan_2}
    \geq \eval{\plan_2}
    \geq \underset{i\in I(A_2)}{\prod} \s_i
    \geq b^{3\underset{i\in I(A_2)}{\sum} a_i}
    \geq b^{3(\frac{S}{2} + 1)}
    \geq b^3\cdot b^{\frac{3}{2}S}
    \geq 8\cdot b^{\frac{3}{2}S} > \beta.
\end{align*}
Therefore, if there are $A_1$ and $A_2$ satisfying \partition problem, then we have a strategy $\plan$ satisfying $\costout{\plan}\leq \beta$. However, if the sets $A_1, A_2$ do not sum to equal value, then any $\plan$ will satisfies $\costout{\plan} > \beta$. Thus, \partition reduces to \mutliplicationOrdering with only powers of $2$. This concludes the proof.
\end{proof}

\section{Future work}\label{section:futurework}

We believe that the bound of \Cref{lemma:product partition is 2-APX} and of \Cref{lemma:notBeing2approwOfPPImpliesNotBeingOPT} can be improved. We would also like to show that the plan that follows a \ppb optimal solution at each level achieves a better approximation ratio. Another direction is to address the remaining cases from  \Cref{prop:optimal stratgy for power of c}, when the single basis is 2 or 3.

Moreover, the \mutliplicationOrdering problem is a specific case of the \textsc{JoinOrdering} problem. Therefore, a future direction is to tackle the case of non-Cartesian product. Namely, instead of having as input only the sizes of the relations $\instance = \{\s_1,\dots,\s_n\}$, the input is also composed of a list of selectivities $\{f_{i,j}\}_{1\leq i,j\leq n}$ that determine the sizes of the joins between relation $i$ and $j$ divided by the size of the Cartesian product between relation $i$ and $j$. Then the only thing that changes in the problem is the definition of the evaluation of a plan $\plan=\plan_1\times\plan_2$ on $\instance$ following $X$ that become $\eval{\plan} = \left(\underset{i\in I(X)}{\prod} \underset{j\in [n]\backslash I(X)}{\prod} f_{i,j}\right) \eval{\plan_1}\cdot\eval{\plan_2}$, where $I(X)$ denotes the set of indices of the elements in $X$. However, this simple change makes the problem much harder as the evaluation is not forced to grow with successive multiplications. This generalization was already partially studied in \cite{DBLP:conf/pods/ChatterjiEGY02}, but one could use a different analysis to tackle special cases or specific conditions.

Lastly, one can consider other cost functions known in the literature and/or other abstract operations (in place of multiplication).
What are the conditions for our results to still hold?

\bibliographystyle{alpha}
\bibliography{biblio}

@article{DBLP:journals/tods/IbarakiK84,
  author       = {Toshihide Ibaraki and
                  Tiko Kameda},
  title        = {On the Optimal Nesting Order for Computing N-Relational Joins},
  journal      = {{ACM} Trans. Database Syst.},
  volume       = {9},
  number       = {3},
  pages        = {482--502},
  year         = {1984},
  url          = {https://doi.org/10.1145/1270.1498},
  doi          = {10.1145/1270.1498},
  bibsource    = {dblp computer science bibliography, https://dblp.org}
}

@article{DBLP:journals/eor/NgBCK10,
  author       = {C. T. Ng and
                  M. S. Barketau and
                  T. C. Edwin Cheng and
                  Mikhail Y. Kovalyov},
  title        = {Product Partition and related problems of scheduling and systems
                  reliability: {C}omputational complexity and approximation},
  journal      = {Eur. J. Oper. Res.},
  volume       = {207},
  number       = {2},
  pages        = {601--604},
  year         = {2010},
  url          = {https://doi.org/10.1016/j.ejor.2010.05.034},
  doi          = {10.1016/J.EJOR.2010.05.034},
  bibsource    = {dblp computer science bibliography, https://dblp.org}
}

@inproceedings{DBLP:conf/coco/Karp72,
  author       = {Richard M. Karp},
  title        = {Reducibility Among Combinatorial Problems},
  booktitle    = {Complexity of Computer Computations},
  pages        = {85--104},
  publisher    = {Plenum Press, New York},
  year         = {1972}
}

@inproceedings{DBLP:conf/icdt/CluetM95,
  author       = {Sophie Cluet and
                  Guido Moerkotte},
  editor       = {Georg Gottlob and
                  Moshe Y. Vardi},
  title        = {On the Complexity of Generating Optimal Left-Deep Processing Trees
                  with Cross Products},
  booktitle    = {Proceedings of the 5th Annual International Conference on Database Theory (ICDT)},
  volume       = {893},
  pages        = {54--67},
  year         = {1995}
}

@inproceedings{DBLP:conf/pods/ChatterjiEGY02,
  author       = {Sourav Chatterji and
                  Sai Surya Kiran Evani and
                  Sumit Ganguly and
                  Mahesh Datt Yemmanuru},
  title        = {On the Complexity of Approximate Query Optimization},
  booktitle    = {Proceedings of the Twenty-first {ACM} {SIGACT-SIGMOD-SIGART} Symposium
                  on Principles of Database Systems (PODS)},
  pages        = {282--292},
  year         = {2002}
}

@inproceedings{DBLP:conf/pods/ScheufeleM97,
  author       = {Wolfgang Scheufele and
                  Guido Moerkotte},
  title        = {On the Complexity of Generating Optimal Plans with Cross Products},
  booktitle    = {Proceedings of the Sixteenth {ACM} {SIGACT-SIGMOD-SIGART} Symposium
                  on Principles of Database Systems (PODS)},
  pages        = {238--248},
  year         = {1997}
}

@online{moerkotte2006building,
  title={Building query compilers (manuscript)},
  author={Moerkotte, Guido},
  url={https://pi3.informatik.uni-mannheim.de/~moer/querycompiler.pdf},
  year={2006}
}

@book{DBLP:books/daglib/0017733,
  author       = {Sanjoy Dasgupta and
                  Christos H. Papadimitriou and
                  Umesh V. Vazirani},
  title        = {Algorithms},
  publisher    = {McGraw-Hill},
  year         = {2008},
  isbn         = {978-0-07-352340-8},
  bibsource    = {dblp computer science bibliography, https://dblp.org}
}

@article{DBLP:journals/corr/abs-2506-17521,
  author       = {Matthias Bentert and
                  Alex Crane and
                  P{\aa}l Gr{\o}n{\aa}s Drange and
                  Yosuke Mizutani and
                  Blair D. Sullivan},
  title        = {Structural Optimal {J}acobian Accumulation and Minimum Edge
                  Count are {\NP}-Complete Under Vertex Elimination},
  journal      = {CoRR},
  volume       = {abs/2506.17521},
  year         = {2025},
  url          = {https://arxiv.org/abs/2506.17521},
  eprinttype   = {arXiv},
  eprint       = {2506.17521}
}

@inproceedings{abo2021approximation,
  title={An Approximation Algorithm for the Matrix Tree Multiplication Problem},
  author={Abo-Khamis, Mahmoud and Curtin, Ryan and Im, Sungjin and Moseley, Benjamin and Ngo, Hung and Pruhs, Kirk and Samadian, Alireza},
  booktitle={46th International Symposium on Mathematical Foundations of Computer Science (MFCS)},
  year={2021},
  organization={Schloss Dagstuhl-Leibniz-Zentrum f{\"u}r Informatik}
}

\end{document}